\documentclass[runningheads]{llncs}
\usepackage[T1]{fontenc}
\usepackage{graphicx}
\usepackage{booktabs}
\usepackage[misc]{ifsym}
\usepackage{amsmath}
\usepackage{amssymb}
\newtheorem{myDef}{Definition}

\usepackage{cite}
\usepackage{multirow}
\usepackage{hyperref}
\usepackage{subcaption}
\usepackage{graphicx}
\usepackage{lipsum}
\newcommand{\corr}{(\Letter)}

\usepackage{mwe}

\begin{document}

\title{\underline{Y}ou \underline{O}nly \underline{T}rain \underline{O}nce: A Flexible Training Framework for Code Vulnerability Detection Driven by Vul-Vector}

\titlerunning{\underline{Y}ou \underline{O}nly \underline{T}rain \underline{O}nce}

\author{Bowen Tian\thanks{The first three authors have contributed equally}\inst{1} \and
Zhengyang Xu\inst{2} \and
Mingqiang Wu\inst{3} \and
Songning Lai\inst{4} \and
Yutao Yue\inst{5} \corr
}

\authorrunning{Bowen Tian et al.}

\institute{
Hong Kong University of Science and Technology (Guangzhou), Deep Interdisciplinary Intelligence Lab \email{bowentian@hkust-gz.edu.cn}
\and
China University of Mining and Technology, Beijing \email{x3542049703@163.com}
\and
China University of Mining and Technology, Beijing \email{mingqiangwu2024@163.com}
\and
Hong Kong University of Science and Technology (Guangzhou), Deep Interdisciplinary Intelligence Lab \email{songninglai@hkust-gz.edu.cn}
\and
Hong Kong University of Science and Technology (Guangzhou), Deep Interdisciplinary Intelligence Lab, Institute of Deep Perception Technology, JITRI, \email{yutaoyue@hkust-gz.edu.cn}
}

\maketitle              

\begin{abstract}
With the pervasive integration of computer applications across industries, the presence of vulnerabilities within code bases poses significant risks. The diversity of software ecosystems coupled with the intricate nature of modern software engineering has led to a shift from manual code vulnerability identification towards the adoption of automated tools. Among these, deep learning-based approaches have risen to prominence due to their superior accuracy; however, these methodologies encounter several obstacles. Primarily, they necessitate extensive labeled datasets and prolonged training periods, and given the rapid emergence of new vulnerabilities, the frequent retraining of models becomes a resource-intensive endeavor, thereby limiting their applicability in cutting-edge scenarios.
To mitigate these challenges, this paper introduces the \underline{\textbf{YOTO}}--\underline{\textbf{Y}}ou \underline{\textbf{O}}nly \underline{\textbf{T}}rain \underline{\textbf{O}}nce framework. This innovative approach facilitates the integration of multiple types of vulnerability detection models via parameter fusion, eliminating the need for joint training. Consequently, YOTO enables swift adaptation to newly discovered vulnerabilities, significantly reducing both the time and computational resources required for model updates. 

\end{abstract}

\section{Introduction}
The extensive application and continuous evolution of computer technology have placed code at the heart of software development, highlighting the critical importance of security issues within this domain\cite{vulnerabilities2005common}. Potential vulnerabilities in code not only compromise individual systems but also jeopardize the overall security and stability of entire industries. Given the vast diversity of software and the increasing complexity of modern software engineering practices, traditional methods of manual code vulnerability identification, which are heavily reliant on human expertise, have become inefficient and costly. As a result, there is a growing trend towards the adoption of automated vulnerability detection tools.\cite{russell2018automated}\cite{harer2018automated}\cite{boudjema2020vyper}

In recent years, deep learning technologies \cite{bowen2024beyond, tian2024pepl, tianimproving} have gained prominence in automated vulnerability detection due to their exceptional accuracy\cite{wang2020combining}\cite{sun2021vdsimilar}\cite{lin2020software}, driven by their ability to model complex patterns within large datasets. However, despite their effectiveness, deep learning-based vulnerability detection methods face substantial challenges\cite{chakraborty2021deep}. Firstly, the training of these models requires vast amounts of labeled data, and the training process itself is computationally intensive and time-consuming. Secondly, the landscape of code vulnerabilities is constantly evolving, with new types of vulnerabilities surfacing regularly. Reintegrating all previously known vulnerability data each time a new type is identified to retrain the model would demand considerable computational resources and time, making the deployment of deep learning-based models in practical applications particularly challenging, especially in scenarios that demand rapid responses to emerging vulnerabilities.

To address these challenges, this paper proposes an innovative method named \underline{\textbf{YOTO}}--\underline{\textbf{Y}}ou \underline{\textbf{O}}nly \underline{\textbf{T}}rain \underline{\textbf{O}}nce framework. By leveraging parameter fusion technology, YOTO integrates models trained on distinct vulnerability datasets without requiring joint training. This approach enables the construction of a deep learning model capable of detecting multiple types of vulnerabilities concurrently. Notably, YOTO can rapidly adapt to the detection of new vulnerabilities while significantly reducing the need for additional training time and computational resources. This breakthrough provides robust support for the practical application of deep learning models in code vulnerability detection, facilitating their deployment in dynamic environments where quick adaptation is paramount.

Specifically, the YOTO method initializes with a pre-trained code feature extraction network and adds classification heads for fine-tuning. Drawing inspiration from the Task Vector methodology \cite{ilharco2022editing}, we capture the direction of parameter movement during the fine-tuning phase for each vulnerability category by hierarchically computing the differences in parameters based on the fine-tuned network weights. This computed vector is termed the Vul-Vector. Subsequently, leveraging the Vul-Vectors derived from multiple vulnerability categories along with the pre-trained weights, a fused model capable of detecting a variety of vulnerability types is synthesized.

Through extensive experimental testing and validation, our YOTO method successfully achieves the effective fusion of multiple common vulnerability types. Under the condition that each vulnerability type is trained independently, our approach yields a model with multi-category vulnerability detection capabilities without necessitating additional training steps. This achievement holds significant theoretical and practical implications for enhancing the adaptability of deep learning-based code vulnerability detection tools to meet the demands of cutting-edge vulnerability detection scenarios.

To better evaluate the effectiveness of the YOTO method, we modified the DiverseVul dataset \cite{chen2023diversevulnewvulnerablesource} using various techniques to simulate different scenarios. Additionally, we utilized CodeT5 \cite{wang2021codet5identifierawareunifiedpretrained}, pre-trained on a large corpus of code texts, as the feature extraction network to ensure proximity to real-world usage conditions. In summary, the main contributions of this paper are outlined below:

\begin{itemize}
    \item Our research introduces a novel Training-Free parameter fusion method into the field of deep learning for code vulnerability detection, marking it as the first work of its kind.
    \item The YOTO method is designed to swiftly adapt to the detection requirements of new vulnerability types without requiring any additional training processes. This characteristic significantly reduces training time and computational resource consumption, thereby facilitating the practical deployment and usability of deep learning-based code vulnerability detection models.
    \item Without the need for joint training, the merged model retains multi-category detection capabilities through parameter fusion alone. Furthermore, the detection performance of the merged model is comparable to that achieved through joint training.
    \item To enhance the community’s understanding of this field and support future developments, we have released the source code in an open-source repository, enabling subsequent researchers to build upon our work.
\end{itemize}

\section{Related Work}

\subsection{Deep Code Vulnerability Detection}
Due to the rapid proliferation in the types and volume of code vulnerabilities, combined with the intricate logic present in engineering code, manual inspection of code vulnerabilities presents a formidable challenge. Traditional rule-based automated vulnerability detection methods, such as those described in \cite{shar2013predicting}, have demonstrated limited efficacy and are impractical for widespread implementation. With the advancement of machine learning and deep learning technologies, systems based on these methodologies are now considered the most promising research direction for code vulnerability detection. For instance, \cite{morrison2015challenges} employs the binary representation of code for vulnerability detection, while \cite{wang2016automatically} leverages deep belief networks to extract semantic features from the abstract syntax trees (ASTs) of Java source code. However, these approaches focus primarily on detection at the source code file level.

Vuldeepecker \cite{li2018vuldeepecker} pioneered the concept of Code Gadgets, which involves extracting API calls from source code files and identifying related statements to integrate into Code Gadgets. These gadgets are then embedded using the Word2Vec model \cite{mikolov2013efficientestimationwordrepresentations}, followed by the training of a code vulnerability detection model utilizing Bi-LSTM. SySeVR \cite{li2021sysevr} presents a deep learning framework for C/C++ code vulnerability detection that operates independently of specific models and categorizes vulnerabilities into syntactic and semantic types. Devign \cite{zhou2019devign} introduces the application of graph neural networks for code feature extraction and vulnerability detection.

In recent years, the development of pre-trained models across various domains has led to fine-tuning from pre-trained models becoming a prevalent training paradigm. Pre-training works like CodeBERT \cite{feng2020codebertpretrainedmodelprogramming} and CodeT5 \cite{wang2021codet5identifierawareunifiedpretrained} have emerged in the field of code representation learning. The enhanced representation capabilities afforded by pre-training have resulted in markedly improved performance in code vulnerability detection tasks compared to earlier models. Building upon the practical advantages of pre-trained models, our work emphasizes the fine-tuning of pre-trained code text models to enhance detection accuracy and efficiency.

\subsection{Parameter Fusion}
The concept of directly fusing model parameters to accomplish certain tasks can be traced back to early federated learning strategies, such as FedAvg \cite{mcmahan2023communicationefficientlearningdeepnetworks}. FedAvg achieves a better centralized model parameter set by calculating the weighted average of client parameters during the training process. However, this fusion approach must be incrementally applied throughout the training phase and has shown marginal improvements in many tasks.

Subsequent research, such as Git Re-basin \cite{ainsworth2209git}, suggests that directly averaging model parameters can lead to performance degradation due to the permutation invariance of neurons in neural networks. Therefore, it recommends aligning neurons in each layer prior to fusion to enhance the effectiveness of parameter aggregation. Nonetheless, studies such as \cite{frankle2020linear, wortsman2022robust, matena2022merging, wortsman2022model} indicate that for models fine-tuned from the same pre-trained weights, sharing a common optimization path (i.e., pre-training) allows direct linear operations to often yield satisfactory results without the need for neuron alignment.

Building on this perspective, Task Vector \cite{ilharco2022editing} proposes editing model parameters by calculating task vectors between fine-tuned parameters and pre-trained parameters and performing vector operations to control the model's ability to merge, forget, and analogize multiple tasks. Inspired by Task Vector, we introduce the idea of calculating fine-tuned model parameter vectors into the field of code vulnerability detection to achieve efficient and accurate model parameter fusion.

\section{Preliminary}

\begin{figure*}[htbp]
\centering
\includegraphics[width=\linewidth]{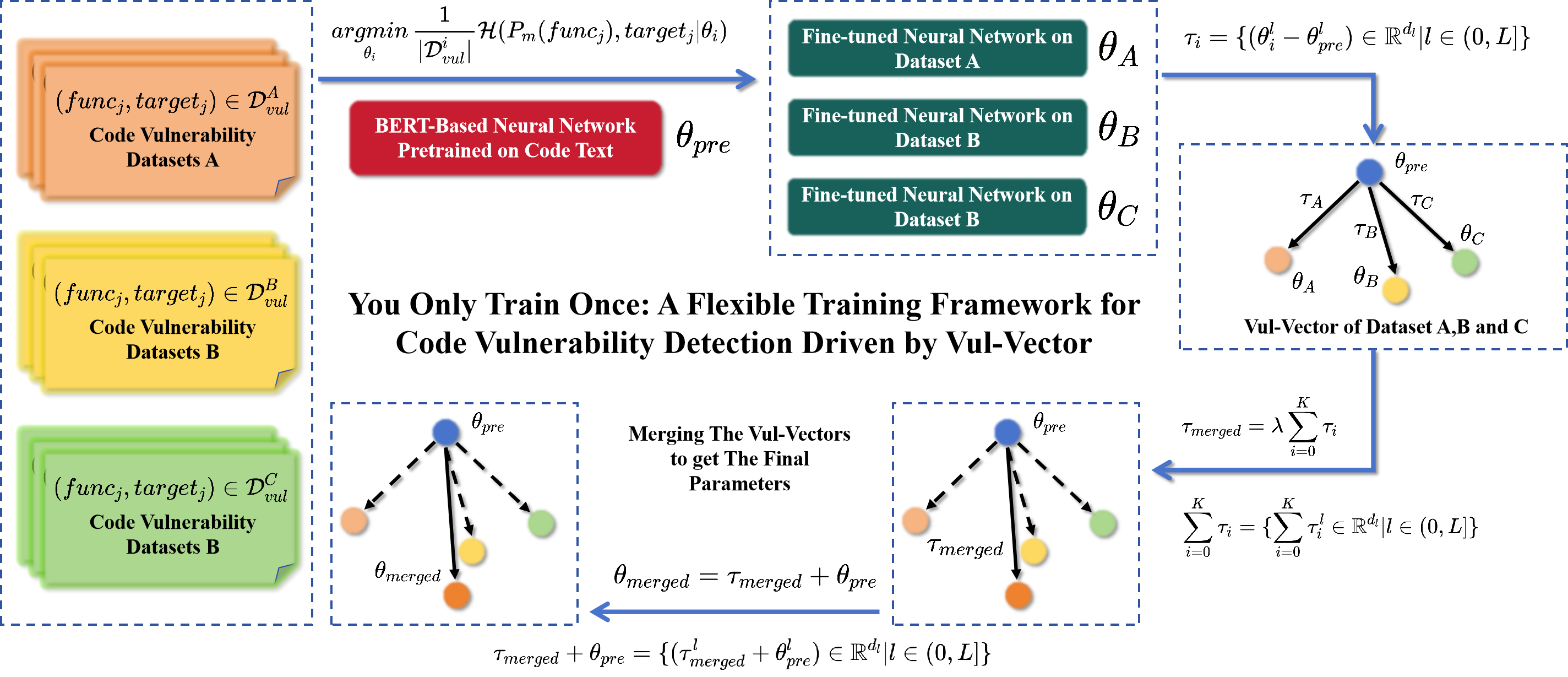}
\caption{An overview of the YOTO training framework.}
\label{figure1}
\end{figure*}
To better illustrate the problem addressed in this article, let's assume that there is a dataset of function-level code vulnerabilities:
\begin{equation}
    \mathcal{D}_{vul} = \{\mathcal{D}^i_{vul}|i\in (0,C]\}
\end{equation}
\begin{equation}
    \mathcal{D}_{vul}^i = \{(func_j,target_j)|j\in N_i\}
\end{equation}
where $D^i_{vul}$ indicates the code vulnerability dataset of a specific vulnerability type (eg. CWE-119), and $C$ denotes the number of vulnerability datasets, $func_j$ represents a function-level piece of code text, while $target_j$ indicates whether the current snippet contains vulnerabilities. In this article, we generally default to the classification of code vulnerabilities following Common Weakness Enumeration (CWE) \cite{vulnerabilities2005common} classification rules.

Traditional methods tend to combine datasets of multiple vulnerability types to form a large dataset, which can be expressed as:
\begin{equation}
\begin{aligned}
\mathcal{D}_{total} &= Concat([\mathcal{D}_{vul}^i],i\in (0,N])\\
&=\{(func_j,target_j)|j\in N_{total}\}
\end{aligned}
\end{equation}
where $N$ represents the number of all samples after merging datasets of different vulnerability categories. The $target_j$ here contains more than just the $0,1$ category, and each vulnerability that is merged into the total dataset is numbered.
Suppose the target model can be expressed as $P_m$, and the cross-entropy loss function can be expressed as $\mathcal{H}(\cdot)$. The training process of the current deep code vulnerability detection model can be expressed as:
\begin{equation}
    \mathop{argmin}\limits_{\theta} \frac{1}{N_{total}}\mathcal{H}(P_m(func_j),target_j|\theta)
\end{equation}
The trained model can classify multiple vulnerabilities.

When new types of vulnerabilities continuously emerge, the aforementioned traditional training process requires us to integrate data on newly discovered vulnerability types into the original training set and retrain the model. This process is obviously extremely computationally intensive. Especially in the context of rapidly evolving vulnerability types, such a training method has become a significant bottleneck restricting the deployment of deep learning-based code vulnerability detection systems in practical applications.

In light of this limitation, we are inspired to explore novel training paradigms within this field: how to enable the model to recognize new types of vulnerabilities while minimizing the need for frequent retraining.

Recent developments in the field of parameter merging give us hope for solving this problem.

Combining the parameters of models with the same architecture trained on different types of datasets to enable the hybrid model to possess the capabilities of all fused models (albeit not as strongly as originally) is the latest hot topic in academia. This idea eliminates the need for joint training, which was previously required to obtain a multitask model.

Model Parameters Combination can generally be expressed as:
\begin{equation}
    \theta_{merged} = \mathcal{M}(\theta_A,\theta_B)
\end{equation}
where $\mathcal{M}$ denotes the function of merging parameters. $\theta_A$ and $\theta_B$ represent the parameters of the two models that are merged, and $\theta_{merged}$ represents the parameters of the merged model.

The early model parameter merging directly averaged the parameters of multiple models to obtain a fusion model, and its calculation process can be expressed as:
\begin{equation}
    \theta_{merged} = \{ \frac{1}{N_m} \sum_{i=0}^{N_m} \theta^l_i | l\in(0,L]\}
\end{equation}
where $N_m$ represents the number of models to be merged and $L$ represents the number of layers of the model.

However, directly averaging the parameters of models with different initialization conditions and training sequences is significantly affected by the permutation invariance of the models (i.e., swapping the positions of any two neurons and their subsequent connections results in an equivalent model), leading to poor performance of the merged model.

To address this issue, Git Re-basin proposes learning a permutation matrix for the model parameters before fusion. This permutation matrix is used to align neurons with the same functionality between the two models. The learning objective can be represented as:
\begin{equation}
    \mathop{argmin}\limits_\pi \parallel \theta_A - \pi(\theta_B) \parallel^2
\end{equation}
where $\pi$ represents the permutation of the model parameters, $\theta_A$ and $\theta_B$ represent the two model parameters that are ready to be aligned.

However, this permutation method is often only suitable for models trained on datasets of the same type, while for multitask learning, the functional relevance of most neurons is relatively weak.

\section{YOTO}

Inspired by works such as \cite{ilharco2022editing}, which have demonstrated the feasibility of linearly interpolating models fine-tuned from the same pre-trained model and achieved promising results in multi-task learning, and noting a certain similarity between the process of merging different code vulnerability detection models and that of multi-task integration, we employ a methodology based on parameter vector algebra to merge code vulnerability detection models trained on different datasets. This approach facilitates the rapid scaling of code vulnerability detection models.

\subsection{Fine-tuning Model Independently}

Unlike the original training paradigm where all types of code vulnerabilities are combined for training, to ensure flexibility, we independently train different vulnerability types during the training process. Each type of vulnerability corresponds to a fine-tuning weight. The training process can be represented as:
\begin{equation}
    \mathop{argmin}\limits_{\theta_i} \frac{1}{|\mathcal{D}_{vul}^i|}\mathcal{H}(P_m(func_j),target_j|\theta_i)
\end{equation}
where $i\in (0,N]$ and $(func_j,target_j) \in \mathcal{D}_{vul}^i$. 

\subsection{Calculating Vul-Vector}

We call the model parameter vectors calculated by using the parameters of the fine-tuned model and the pre-trained model Vul-Vector, Its calculation process can be simply represented as follows:

\begin{equation}
     \tau_i = \{ (\theta_i^l - \theta_{pre}^l) \in \mathbb{R}^{d_l} | l\in(0,L]\}
\end{equation}

where $\mathbb{R}^{d_l}$ represents the space in which the dimension of the $l$ layer is located, $\tau_i$ represents the Vul-Vector of the fine-tuned model $\theta_i$.

\subsection{Assembling Vul-Vectors}

After obtaining multiple Vul-Vectors for different types of vulnerability datasets, based on the assumption that fine-tuned weights are located in the same flat loss basin \cite{li2023deep}, we perform vector addition on these Vul-Vectors through a linear combination. This results in a comprehensive parameter vector that integrates multiple Vul-Vectors. We believe this parameter vector consolidates the directions of parameter changes when the pre-trained model weights are fine-tuned on different types of vulnerability datasets.

To accomplish the addition of Vul-Vector, we can define a parametric operation:

\begin{myDef}
\label{def:1}
Any two Vul-Vector obtained from the same pre-trained model, their addition operations can be defined as:

\begin{equation}
    \tau_n + \tau_m = \{(\tau^l_n + \tau^l_m) \in \mathbb{R}^{d_l} |l\in (0,L]\}
\end{equation}

Satisfy the commutative and associative laws.
\end{myDef}

Therefore, the addition between Vul-Vector can be easily generalized to the continuous addition scenario:

\begin{myDef}
\label{def:definition2}
The addition of $K$ Vul-Vectors can be expressed as:
\begin{equation}
    \sum\limits_{i=0}^K \tau_i =  \{\sum\limits_{i=0}^K \tau_i^l \in \mathbb{R}^{d_l} |l\in (0,L]\}
\end{equation}
\end{myDef}

When we integrate multiple Vul-Vectors, we can directly add them according to \autoref{def:definition2}, and finally multiply a coefficient lambda that controls the size of the merged Vul-Vector to get the final integrated vector:
\begin{equation}
    \tau_{merged} = \lambda \sum\limits_{i=0}^K \tau_i
\end{equation}
where $\lambda$ is given in the form of hyperparameters.

\subsection{Application of Vul-Vector}

The $\tau_{merged}$ obtained in the previous step only integrates the weight movement directions when the pre-trained model theta is fine-tuned on multiple different vulnerability datasets. Therefore, to obtain an integrated model with detection capabilities across multiple vulnerability datasets, it is necessary to fuse $\tau_{merged}$ with the original pre-trained model weights $\theta_{pre}$ to derive the final model weights. The calculation process can be easily represented as:

\begin{equation}
    \tau_{merged} + \theta_{pre} = \{(\tau^l_{merged} + \theta^l_{pre}) \in \mathbb{R}^{d_l} |l\in (0,L]\}
\end{equation}

We can record the final weights as $\theta_{merged} = \tau_{merged} + \theta_{pre}$. $\theta_{merged}$ is the final output of the YOTO method, which is considered to be a fusion of multiple vulnerability detection capabilities.

\section{Experiments}
In this section, we will introduce our experimental setup and results in detail to prove the effectiveness of YOTO.

\subsection{Dataset}
The \textbf{DiverseVul} \cite{chen2023diversevulnewvulnerablesource} dataset extracts 330,492 complete functions from 797 large C/C++ projects, encompassing 150 common CWE vulnerabilities. Among these, 16,109 vulnerable functions have CWE type information. We also extended our experiments to two other C/C++ function-level vulnerability datasets, \textbf{Draper} \cite{boudjema2020vyper} as well as \textbf{Bigvul} \cite{fan2020ac}. In order to explore the effectiveness of our approach on other programming languages, we also added the \textbf{Reposvul} dataset \cite{wang2024repository} (containing different languages such as C/C++, Python, and Java).

\subsection{Training Setup}
To ensure that the final experimental results are not influenced by the training process, we adopt the same training procedure for different dataset construction methods. We utilize CodeT5-base (BERT-based) as the network for code feature extraction. CodeT5 has been pre-trained on a large corpus of code text for various tasks and can perform code generation or code understanding tasks. We only use the Encoder layer of CodeT5 to extract function-level code features (hereinafter referred to as the feature extraction layer). Subsequently, we construct a classification head consisting of fully connected layers based on the number of categories in the current training data. For the feature extraction layer with pre-trained parameters, we fine-tune it using a learning rate of 1e-5, while for the classification head, we optimize its parameters using a learning rate of 1e-2. We train all datasets for 128 epochs with a batch size of 16 and use the Adam optimization algorithm for gradient descent.In order to alleviate the common data imbalance in the code vulnerability dataset, we use the weighted cross-entropy loss function to learn the classifier, and the loss weight of all vulnerability classifications is $5$ times that of the non-vulnerability.

\subsection{An Evaluation of Single Vulnerability Classification}

\begin{table}[htbp]
  \begin{center}
    \caption{Information About the Constructed Dataset}
    \label{tab:table1}
    \resizebox{0.6\columnwidth}{!}{
    \begin{tabular}{ccc}
      \toprule[0.5mm]
      \textbf{Vulnerability Type} & \textbf{Num of Vul Func} & \textbf{Total Num}\\
      \midrule
      \textbf{CWE-190} & 674 & 9760 \\
      \textbf{CWE-617} & 162 & 3492 \\
      \textbf{CWE-772} & 100 & 1439 \\
      \textbf{CWE-269} & 111 & 1839 \\
      \bottomrule[0.5mm]
    \end{tabular}
    }
  \end{center}
\end{table}

\noindent \textbf{Dataset Construction}: Firstly, we segment the dataset with the finest granularity by dividing the completed DiverseVul dataset according to CWE types. Each subset contains only one type of CWE, and the data within it is categorized as either having vulnerabilities of that CWE type or not having vulnerabilities of that CWE type. Through this segmentation method, we divide DiverseVul into 150 subsets. To facilitate the evaluation of our method, we randomly sample four datasets from these subsets, which cover different data distributions. Their specific details are shown in \autoref{tab:table1}.

\noindent \textbf{Directly Fine-tuning}: We train on these datasets independently to create four binary classifiers that are independent of each other. The experimental results are shown in \autoref{tab:table2}.

\begin{table}[h]
  \begin{center}
    \caption{Directly Fine-tuning for Single Vulnerability}
    \label{tab:table2}
    \resizebox{0.6\columnwidth}{!}{
    \begin{tabular}{cccc}
      \toprule[0.5mm]
      \textbf{Vulnerability Type} & \textbf{Accuracy} & \textbf{Recall} & \textbf{Precision} \\
      \midrule
      \textbf{CWE-190} & 86.91 & 17.47 & \underline{20.83} \\
      \textbf{CWE-617} & \underline{91.12} & 17.24 & 19.14 \\
      \textbf{CWE-772} & 86.11 & 14.81 & 19.44 \\
      \textbf{CWE-269} & 90.09 & \underline{18.30} & 18.48 \\
      \bottomrule[0.5mm]
    \end{tabular}
    }
  \end{center}
\end{table}

Due to the small amount of data in general, the vulnerability patterns identified by the model are not comprehensive enough, so Recall and Precision are low, but compared with common deep code vulnerability detection models\cite{chen2023diversevulnewvulnerablesource}, this result is within a reasonable range in single classification.

\begin{table}[h]
  \begin{center}
    \caption{Merging Test for Multi-Vulnerability}
    \label{tab:table2.1}
    \resizebox{0.6\columnwidth}{!}{
    \begin{tabular}{ccccc}
      \toprule[0.5mm]
      \textbf{Model} & \textbf{Dataset Index} & \textbf{Accuracy} & \textbf{Recall} & \textbf{Precision} \\
      \midrule
      \multirow{3}{*}{CWE-772} & CWE-269 & 55.3 & 8.24 & 3.91 \\
      ~ & CWE-617 & 92.63 & 7.71 & 7.14 \\
      ~ & CWE-190 & 78.89 & 14.62 & 12.23 \\
      \midrule
      \multirow{3}{*}{CWE-269} & CWE-772 & 54.17 & 11.33 & 10.94 \\
      ~ & CWE-617 & 85.76 & 14.76 & 18.00 \\
      ~ & CWE-190 & 84.04 & 12.59 & 16.22 \\
      \midrule
      \multirow{3}{*}{CWE-617} & CWE-772 & 76.39 & 16.44 & 19.17 \\
      ~ & CWE-269 & 65.22 & 12.75 & 27.17 \\
      ~ & CWE-190 & 68.52 & 14.14 & 12.45 \\
      \midrule
      \multirow{3}{*}{CWE-190} & CWE-772 & - & 0 & 0 \\
      ~ & CWE-269 & 61.82 & 2.82 & 7.07 \\
      ~ & CWE-617 & - & 0 & 0\\
      \midrule
      \multirow{4}{*}{Param Mean} & CWE-772 & 80.56 & 16.20 & 26.39 \\
      ~ & CWE-269 & 78.89 & 10.36 & 18.02 \\
      ~ & CWE-617 & - & 0 & 0\\
      ~ & CWE-190 & 80.73 & 13.42 & 15.57 \\
      \midrule
      \multirow{4}{*}{\textbf{YOTO(Ours)}} & CWE-772 & 81.57 & 18.06 & 31.94 \\
      ~ & CWE-269 & 79.62 & 15.4 & 22.83 \\
      ~ & CWE-617 & 90.98 & 13.62 & 15.71 \\
      ~ & CWE-190 & 80.81 & 17.34 & 26.95 \\
      \bottomrule[0.5mm]
    \end{tabular}
    }
  \end{center}
\end{table}

\noindent \textbf{Merging Test}: We integrated four single-classification models using YOTO. Specifically, during the integration process, we only considered the feature extraction layers, while the classification layers were replaced based on the evaluation dataset.The merging coefficient is determined through testing on a validation set randomly sampled from the training set. In this experiments $\lambda$ is set to $0.3$. We designed three types of experiments: the first type was a cross-experiment, which involved directly testing models fine-tuned on one dataset on different datasets; the second type evaluated the directly averaged models; and the third type evaluated the models integrated using the YOTO method. The specific experimental results are shown in \autoref{tab:table2.1}.

As can be seen from the results, the model after fusion using YOTO exhibits remarkable stability across various datasets. Although some precision is sacrificed after fusion, the average performance evaluated on multiple datasets is significantly higher than that of the individual models before fusion. Moreover, it still maintains a strong advantage compared to direct average fusion.

\subsection{An Evaluation of Multi-Vulnerability Classification}
\label{sec:multi}

\begin{figure}
    \centering
    \begin{subfigure}{0.3\linewidth}
        \centering
        \includegraphics[width=\linewidth]{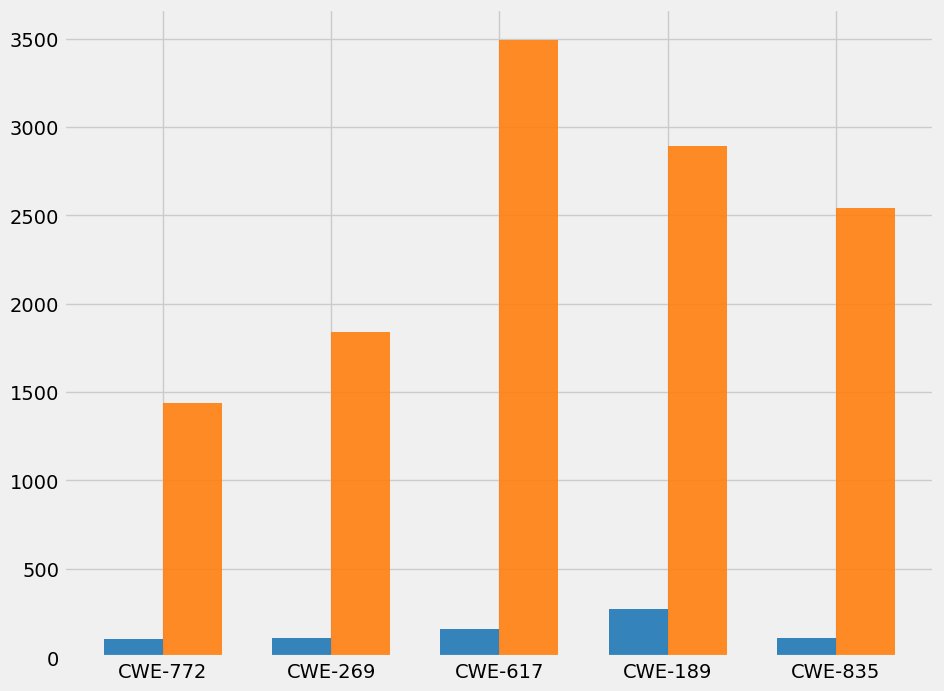}
        \caption{Dataset A}
        \label{fig:part1}
    \end{subfigure}
    \begin{subfigure}{0.3\linewidth}
        \centering
        \includegraphics[width=\linewidth]{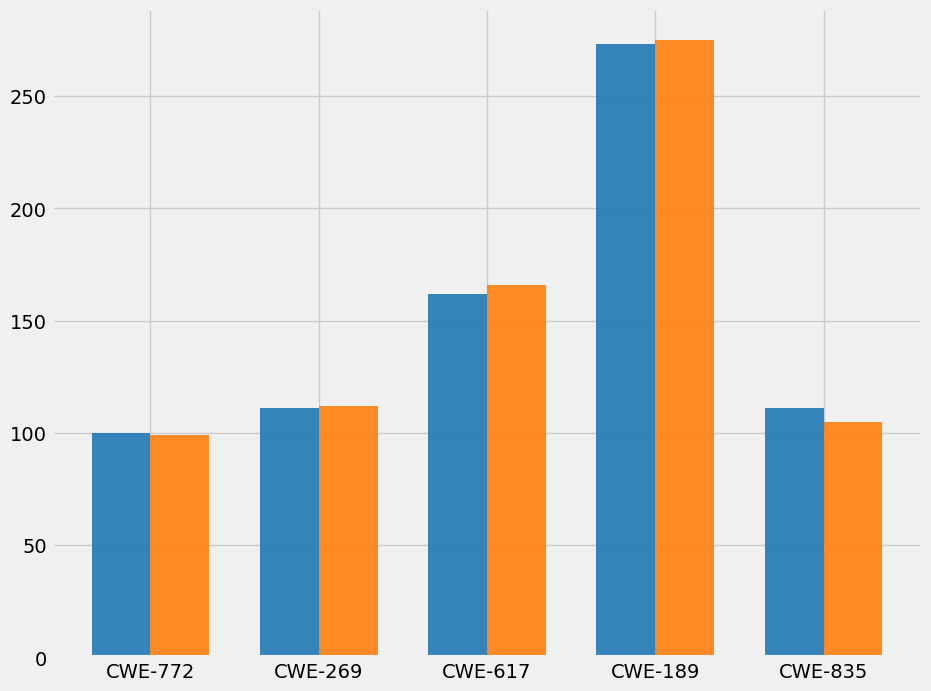}
        \caption{Dataset A Sampled}
        \label{fig:part2}
    \end{subfigure} \\
    \begin{subfigure}{0.3\linewidth}
        \centering
        \includegraphics[width=\linewidth]{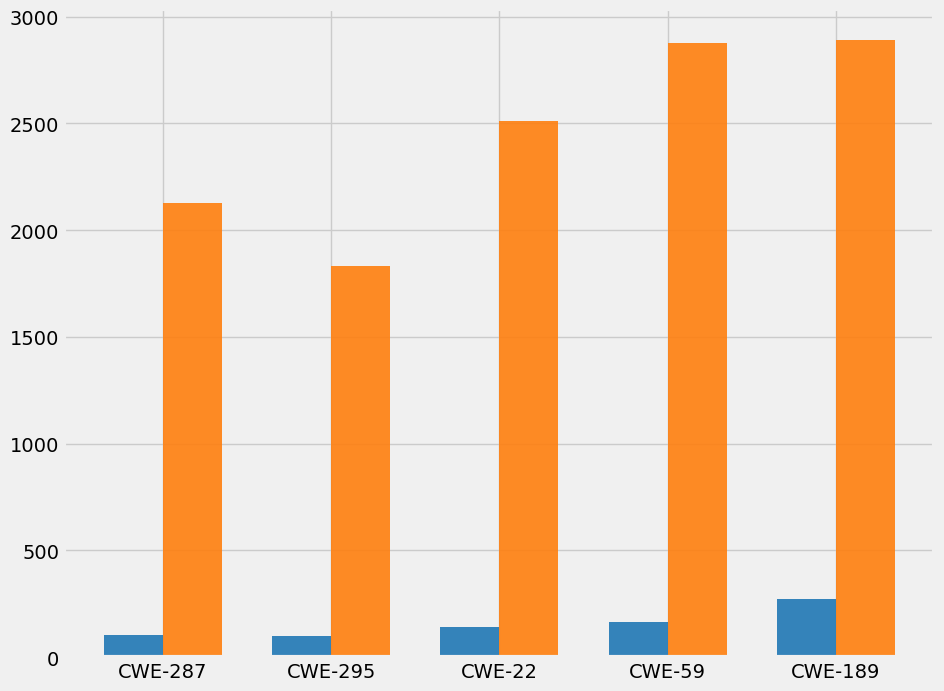}
        \caption{Dataset B}
        \label{fig:part3}
    \end{subfigure}
    \begin{subfigure}{0.3\linewidth}
        \centering
        \includegraphics[width=\linewidth]{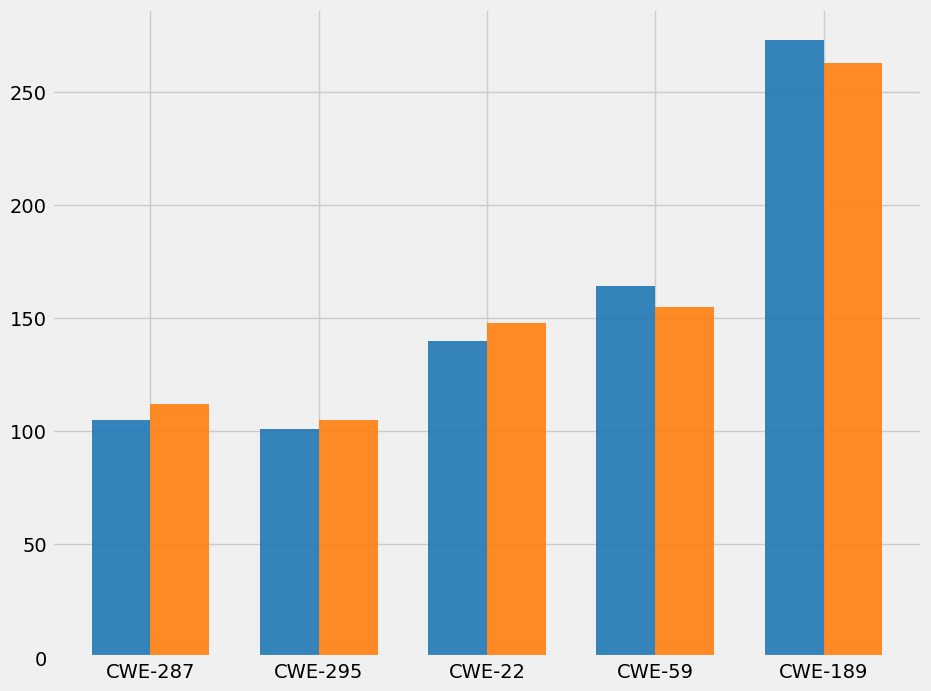}
        \caption{Dataset B Sampled}
        \label{fig:part4}
    \end{subfigure}
    \caption{Data Distribution Before and After Sampling in \autoref{sec:multi}. The left-hand side panels, \autoref{fig:part1} and \autoref{fig:part3}, illustrate the original data distributions of Dataset A and Dataset B, respectively. The right-hand side panels, \autoref{fig:part2} and \autoref{fig:part4}, depict the data distributions of Dataset A and Dataset B after sampling. With a more balanced data distribution, the effectiveness of the method can be better observed.}
    \label{fig:data_dist}
\end{figure}

\noindent \textbf{Dataset Construction}: To further evaluate the effectiveness of the method, we selected 10 CWE vulnerability types and divided them into two sets of five CWEs each, forming two 6-class datasets (one class representing no vulnerability and the other five classes representing five different CWE vulnerabilities). The label distribution and quantities of these ten CWE vulnerability types can be observed in Figure 2. Due to the relatively high number of samples labeled as no vulnerability, in order to better observe the experimental results (from the perspective of assessing method performance), we conducted sampling on the no-vulnerability data to match the size of the vulnerable dataset. The data distribution after sampling is also presented in \autoref{fig:data_dist}.

\begin{table}[htbp]
  \begin{center}
    \caption{Directly Fine-tuning for Multi-Vulnerability}
    \label{tab:table3}
    \resizebox{0.5\columnwidth}{!}{
    \begin{tabular}{cccc}
      \toprule[0.5mm]
      \textbf{Dataset Index} & \textbf{Accuracy} & \textbf{Recall} & \textbf{Precision} \\
      \midrule
      Dataset A & 70.84 & 70.11 & 72.08 \\
      Dataset B & 72.97 & 70.58 & 77.98 \\
      \bottomrule[0.5mm]
    \end{tabular}
    }
  \end{center}
\end{table}

\noindent \textbf{Directly Fine-tuning}: We train on these datasets independently to create two 6-class classifiers that are independent of each other. The experimental results are shown in \autoref{tab:table3}, Due to the increase in the amount of data, the multi-classification model has performed well in various metrics.

\begin{table}[htbp]
  \begin{center}
    \caption{Merging Test for Multi-Vulnerability}
    \label{tab:table4}
    \resizebox{0.6\columnwidth}{!}{
    \begin{tabular}{ccccc}
      \toprule[0.5mm]
      \textbf{Model} & \textbf{Dataset Index} & \textbf{Accuracy} & \textbf{Recall} & \textbf{Precision} \\
      \midrule
      Model A & Dataset B & 44.40 & 65.40 & 54.96 \\
      \midrule
      Model B & Dataset A & 48.22 & 65.00 & 55.77 \\
      \midrule
      \multirow{2}{*}{Param Mean} & Dataset A & 69.84 & 66.71 & 63.66 \\
      ~ & Dataset B & 69.03 & 67.17 & 74.35 \\
      \midrule
      \multirow{2}{*}{YOTO(\textbf{Ours})} & Dataset A & 70.98 & 69.22 & 70.36 \\
      ~ & Dataset B & 70.03 & 69.50 & 80.15 \\
      \bottomrule[0.5mm]
    \end{tabular}
    }
  \end{center}
\end{table}

\noindent \textbf{Merging Test}: Next, we employ the YOTO method to merge the parameters of these two multi-class classification models. In this merging experiment, the merging coefficient $\lambda$ is set to $0.6$. Similarly, we only merge the feature extraction layers. The experimental results are evaluated from three perspectives. The first is cross-evaluation, where the multi-class vulnerability detection model trained on Dataset A is evaluated on the test set of Dataset B. Since several CWE vulnerability patterns share some common characteristics, this type of evaluation yields somewhat effective results. The second perspective involves evaluating the model obtained by directly averaging the parameters. The third perspective involves evaluating the model obtained using the YOTO method. The experimental results are presented in \autoref{tab:table4}.

From the experimental results, it can be observed that after combining the parameters of two multi-class models using the YOTO method, the resulting new model can simultaneously detect vulnerability types in both Dataset A and Dataset B. The performance of this new model is significantly improved compared to direct detection($26.58\%$ in Accuracy, almost $25\%$ in Precision, and $4.5\%$ in Recall). Moreover, it also shows a notable improvement compared to the method of directly mixing parameters. This fully demonstrates the effectiveness of the YOTO method.

\begin{table}[htbp]
\begin{center}
\caption{Experimentation on a wider dataset}
\label{tab:ext}
\resizebox{0.6\columnwidth}{!}{
\begin{tabular}{cccccc}
\toprule[0.5mm]
Dataset                   & Model                       &           & Accuracy & Recall & Precision \\
\midrule
\multirow{6}{*}{Bigvul}   & model A                     & dataset B & 58.14    & 79.67  & 89.55     \\
\cmidrule(lr){2-6}
                          & model B                     & dataset A & 55.48    & 74.76  & 89.57     \\
                          \cmidrule(lr){2-6}
                          & \multirow{2}{*}{Param Mean} & dataste A & 69.01    & 84.24  & 84.39     \\
                          &                             & dataset B & 71.92    & 79.88  & 85.41     \\
                          \cmidrule(lr){2-6}
                          & \multirow{2}{*}{YOTO(Ours)} & dataset A & 76.77    & 83.72  & 92.68     \\
                          &                             & dataset B & 75.88    & 79.57  & 92.67     \\
\midrule
\multirow{6}{*}{Reposvul} & model A                     & dataset B & 50.86    & 48.67  & 71.41     \\
\cmidrule(lr){2-6}
                          & model B                     & dataset A & 43.96    & 64.75  & 63.57     \\
                          \cmidrule(lr){2-6}
                          & \multirow{2}{*}{Param Mean} & dataste A & 78.13    & 81.71  & 79.44     \\
                          &                             & dataset B & 70.82    & 87.87  & 71.57     \\
                          \cmidrule(lr){2-6}
                          & \multirow{2}{*}{YOTO(Ours)} & dataset A & 79.17    & 86.05  & 82.01     \\
                          &                             & dataset B & 73.61    & 84.29  & 78.28     \\
\midrule
\multirow{6}{*}{Draper}   & model A                     & dataset B & 50.75    & 11.2   & 43.57     \\
\cmidrule(lr){2-6}
                          & model B                     & dataset A & 49.87    & 14.78  & 47.14     \\
                          \cmidrule(lr){2-6}
                          & \multirow{2}{*}{Param Mean} & dataste A & 82.22    & 45.73  & 50.02     \\
                          &                             & dataset B & 63.26    & 39.81  & 49.97     \\
                          \cmidrule(lr){2-6}
                          & \multirow{2}{*}{YOTO(Ours)} & dataset A & 88.13    & 46.12  & 50.05     \\
                          &                             & dataset B & 65.95    & 35.11  & 49.99     \\
\bottomrule[0.5mm]
\end{tabular}
}
\end{center}
\end{table}

In order to verify the robustness of YOTO, we modeled the multi-vulnerability dataset construction method described above and conducted experiments on a broader code vulnerability detection dataset. We extended three datasets, the C/C++ function-level vulnerability datasets Bigvul \cite{fan2020ac} and Draper \cite{russell2018automated}, and the multi-programming-language vulnerability dataset Reposvul \cite{wang2024repository} containing different granularities.We still selected a larger number of CWE classifications from these datasets to form two subdatasets to train the model separately, and then used the fusion method to perform parameter fusion and testing, and the results are shown in \autoref{tab:ext}. This result further illustrates the superiority of YOTO.

\subsection{An Evaluation of Incremental Learning}

To better simulate the effectiveness of the YOTO method in practical scenarios, given the existing multi-class classification model and the independently trained single-class classification models, we integrated the parameters of the single-class models into the multi-class model's parameters, thereby enabling the multi-class model to acquire new vulnerability detection capabilities. Specifically, we directly utilized the pre-trained multi-class and single-class models from the previous two sections (notably, for the multi-class model, we chose Model B trained on Dataset B). We iteratively incorporated new single-class models into the multi-class model until all were fully integrated. The evaluation results are presented in \autoref{tab:table5}:

\begin{table}[htbp]
  \begin{center}
    \caption{Results of the incremental learning experiment}
    \label{tab:table5}
    \resizebox{0.6\columnwidth}{!}{
    \begin{tabular}{ccccc}
      \toprule[0.5mm]
      \textbf{Model} & \textbf{Dataset Index} & \textbf{Accuracy} & \textbf{Recall} & \textbf{Precision} \\
      \midrule
      \multirow{2}{*}{Model C} & Dataset B & 69.61 & 68.74 & 74.85 \\
      ~ & CWE-772 & 85.07 & 18.52 & 25 \\
      \midrule
      \multirow{3}{*}{Model D} & Dataset B & 65.83 & 67.26 & 73.24 \\
      ~ & CWE-772 & 80.56 & 15.74 & 26.39 \\
      ~ & CWE-269 & 85.6 & 16.85 & 20.57 \\
      \midrule
      \multirow{4}{*}{Model E} & Dataset B & 65.55 & 67.05 & 72.24 \\
      ~ & CWE-772 & 79.86 & 16.67 & 29.17 \\
      ~ & CWE-269 & 82.47 & 16.21 & 21.74 \\
      ~ & CWE-617 & 90.48 & 12.86 & 13.71 \\
      \midrule
      \multirow{5}{*}{Model F} & Dataset B & 63.45 & 66.06 & 75.08 \\
      ~ & CWE-772 & 79.51 & 16.44 & 29.17 \\
      ~ & CWE-269 & 81.39 & 16.30 & 23.91 \\
      ~ & CWE-617 & 90.48 & 12.57 & 14.00 \\
      ~ & CWE-190 & 79.30 & 17.28 & 28.59 \\
      \bottomrule[0.5mm]
    \end{tabular}
    }
  \end{center}
\end{table}

Model C represents the parameter fusion of Model B and the CWE-772 model using YOTO, while Model D further incorporates the model parameters of CWE-269 based on Model C. Following this pattern, Model F integrates the parameters of Model B and four single-category vulnerability detection models. The experimental results demonstrate that, while maintaining the original model's multi-classification capabilities without significant impact, the model can simultaneously adapt to newly introduced vulnerability types. This experiment further validates the rapid scalability of the YOTO method in real-world scenarios.

\section{Conclusion}
This paper proposes a flexible training paradigm for code vulnerability detection based on the method of direct parameter fusion. Firstly, models with different vulnerability detection capabilities are obtained by fine-tuning a pre-trained model independently on datasets of different vulnerability types. The YOTO method proposed herein utilizes the parameters of these models and the pre-trained model to compute Vul-Vector parameter vectors, which represent the directions of parameter shifts during the fine-tuning of different vulnerability detection models. By merging these Vul-Vectors and applying them to the pre-trained model parameters, a model with multiple vulnerability detection capabilities can be obtained without additional training. This algorithm is expected to facilitate the rapid adaptation of existing code vulnerability detection models to new types of vulnerabilities without the need for joint training, thereby promoting the practical application of deep learning-based code vulnerability detection methods.

\section{Limitations}
While the experimental section of this paper has evaluated the effectiveness of the YOTO method in three common scenarios through different dataset construction methods, the evaluations were conducted using only the DiverseVul dataset and a single feature extraction network for code (CodeT5). Therefore, future work could extend the method to a broader range of datasets and pre-trained networks to obtain more objective evaluation results. Although YOTO achieved promising results in the experiments, methods based on parameter fusion for code vulnerability detection are more difficult to interpret compared to joint training methods. Thus, exploring the interpretability of the YOTO method will also be a valuable research direction.

\section{Acknowledgement}
in part by Guangzhou-HKUST(GZ) Joint Funding Program (Grant No.2023A03J0008) Education Bureau of Guangzhou Municipality.


%
%
%
%

\end{document}